\documentclass[]{aa}
\usepackage{amsmath}
\usepackage{graphicx}
\usepackage{txfonts}
\usepackage{tikz}
\AtBeginDocument{\mathcode`v=\varv}

\begin{document}

   \title{Strong surface outflows on accretion discs}

   \author{C. J. Nixon
          \inst{1}
          \and
          J. E. Pringle\inst{1,2}
          }

   \institute{Department of Physics and Astronomy, University of Leicester, Leicester, LE1 7RH, UK\\
              \email{cjn@leicester.ac.uk}
         \and
             Institute of Astronomy, Madingley Road, Cambridge, CB3 0HA, UK
             }

   \date{\today}


   \abstract{In order to provide an explanation for the unexpected radial brightness distribution of the steady accretion discs seen in nova-like variables, \cite{Nixon:2019ab} proposed that the accretion energy is redistributed outwards by means of strong, magnetically driven, surface flows. In this paper we note that  the ``powerful, rotating disc winds'' observed in the soft states of black hole X-ray binaries, and also in the disc around a magnetized neutron star in Her X-1, have the properties of the outflows postulated by Nixon \& Pringle to exist in the nova-like variable accretion discs around white dwarfs. The relevant properties are that the flows are not winds, but are, instead, bound flows (traveling at less than the escape velocity) and that the mass fluxes in the flows are a substantial fraction of the accretion rate in the disc.}

   \keywords{accretion, accretion disks --- magnetic fields --- magnetohydrodynamics(MHD) --- novae, cataclysmic variables --- black hole physics --- X-rays: binaries}

   \maketitle

\section{Introduction}
\label{intro}

In a recent paper, \cite{Nixon:2019ab} drew attention to the observational evidence that in the (quasi-)steady accretion discs in nova-like variables the radial distribution of emission of radiation does not agree with the expected radial distribution of dissipation of accretion energy. The observations indicate that there is a severe energy deficit in the inner disc regions. This implies that the accretion energy is not radiated away from where it is released. \cite{Nixon:2019ab} speculated that the MHD turbulence, which is thought to power the accretion in such fully-ionized, high-viscosity discs, gives rise to magnetically driven, corona-like structures (which, to avoid confusion they named magnetically controlled zones, or MCZs). They suggested that these give rise to the outward transfer of mass, energy and angular momentum in the form of  a low density, high velocity flow in the low density regions of the disc above and below the disc plane. \cite{Nixon:2019ab} proposed a simple toy model for the flows in the MCZs, and showed that approximate agreement with the observations could be achieved if the mass outflow in the MCZs was in the range of $\mu \approx 10 - 30$  percent of the accretion rate, and if, on average, the effect of the MCZ was to throw material outwards a distance of around $k \approx 1.5 - 3$ times the launching radius. 

In order to estimate the typical radial velocities involved in such a process we note that if material in a disc at radius $R$ around a central star of mass $M$ and circular velocity $v_\phi = (1/\sqrt{2}) v_{\rm esc}$, where the escape velocity
\begin{equation}
v_{\rm esc} =  \left( \frac{2GM}{R} \right)^{1/2},
\end{equation}
is impulsively given, in addition, a radial velocity $v_R = f v_\phi$ (with $f < 1$, i.e. a bound flow), then assuming conservation of angular momentum, the material reaches a radius $kR$ where $k = 1/(1-f)$. Of course, if $f \ge 1$ then the material is unbound, and the outward flow becomes a wind. Thus the values of $f$ required by the toy model of Nixon \& Pringle are in the range $f \approx 0.3 - 0.7$. 

In Section~\ref{Her} we discuss the properties of the ``accretion disc wind'' in the X-ray binary system Her X-1/HZ Her, reported by \cite{Kosec:2020aa}. The geometry of this system relative to the observer is well established, and the line of sight passes close to the disc surface. We find that the flow velocities are mostly below the escape velocity, and conclude that this is not a wind in the conventional sense. However, the flow properties, in terms of velocities and mass flux, agree with those postulated in \cite{Nixon:2019ab}. In Section~\ref{discwinds} we discuss the properties of the equatorial ``disc winds'' seen in the soft states of black hole binaries. Here the geometry is not so exactly established, but circumstantial evidence suggests that the systems are all at high inclination \citep{Ponti:2012aa}. Again we find that these flows are not unbound and so do not correspond to winds. But the flow properties do align with those proposed by \cite{Nixon:2019ab}, both in terms of velocities and in terms of mass fluxes. We present discussion in Section~\ref{conclusions}.

\section{Hercules X-1}
\label{Her}
\cite{Kosec:2020aa} present evidence for outward flow of material close to the disc surface in the X-ray binary Hercules X-1.

Hercules X-1 was one of the brightest and most intriguing X-ray sources discovered by the UHURU satellite \citep{Tananbaum:1972aa}. The X-ray emission displayed regular pulses with a period of 1.24 seconds, regular eclipses with a period of 1.7 days, and a regular cycle of appearances and disappearances with a period of around 35 days. The suggested optical identification with the variable star HZ Her by \cite{Liller:1972aa} was confirmed by the discovery that this star showed a light curve with the same 1.7 day period \citep{Bahcall:1972aa}. 

The short period X-ray pulsations were quickly identified as being due to accretion onto a rotating, strongly magnetized neutron star \citep[][see also \citealt{Pringle:1972aa}]{Davidson:1972aa}. The X-ray eclipses are caused by occultations of the X-ray flux by a binary companion, and the main 1.7 day variation of the optical light curve is identified as primarily caused by heating of one side of the star by the X-ray flux from its compact companion \citep{Bahcall:1972aa,Joss:1973aa,Pringle:1973aa}. 

Thus, Her X-1 is the X-ray counterpart of the variable star system  HZ Her. The binary system consists of a 2.3 $M_\odot$ primary star of spectral type F to A depending on whether one is observing the X-ray illuminated side or not. The secondary (less massive) component is a neutron star with assumed mass 1.4 $M_\odot$. There is Roche lobe mass transfer from the primary to the secondary, forming an accretion disc. The disc is truncated by the magnetosphere of the neutron star at a radius of about $R \approx 2 \times 10^8$\,cm. The outer edge of the disc is at $R \approx 2 \times 10^{11}$\,cm, and the binary separation is around $a \approx 6 \times 10^{11}$\,cm.

 Identification of the cause of the 35 day X-ray light curve proved more problematic. \cite{Gerend:1976aa} proposed a phenomenological model for this behaviour, which also explained subtle variations in the optical light curve over the 35 day cycle. They proposed that the neutron star was encircled by a large, tilted accretion disc, which precessed with the period of 35 days. The explanation as to why the accretion disc might be tilted and precess was later provided in terms of radiation-driven disc warping \citep{Iping:1990aa,Pringle:1996aa,Wijers:1999aa}. This model is also capable of providing an explanation for detailed behaviour of the X-ray flux through the 35 day period \citep[see, for example,][]{Scott:2000aa,Leahy:2002aa}. 

In this system, \cite{Kosec:2020aa} report evidence for blue-shifted absorption features, seen primarily in high ionization lines of N VII, O VIII, N X and Fe XXV/XXVI. For these features they identify (given in their Table~3) a radial velocity ($v_R$), relative to the neutron star, as well as an ionization parameter, $\xi$, for the absorbing plasma. From the ionization parameter they are able to estimate a maximum radial distance, $R_{\rm max}$,  that the absorbing material can have from the neutron star (given in their Table~5). In line with previous work (see Section~\ref{discwinds}) \cite{Kosec:2020aa} interpret this outward moving material as evidence for a disc wind.

As noted by \cite{Kosec:2020aa}, the absorption features seen in the Her X-1 X-ray data lie along the line of sight between the observer and the neutron star.  Moreover, because the 35 day variation in the Her X-1/HZ Her system is caused by a warped and tilted disc moving in and out of the line of sight to the neutron star, it is evident that when the X-ray source is visible, those lines of sight must lie close to the surface of the disc. Thus, in the Nixon \& Pringle model of a steady disc, those lines of sight would be passing through the postulated MCZ.

We now look at the wind hypothesis in more detail. One basic property of a wind is that the velocity of the wind should exceed the escape velocity. For typical winds, the wind velocity exceeds the escape velocity by a factor of a few.  At a given radius, $R$,  the escape velocity is
\begin{equation}
v_{\rm esc} = 1933 \left( \frac{M}{1.4 M_\odot} \right)^{1/2} \left( \frac{R}{10^{10} {\rm cm}} \right)^{-1/2} {\rm km/s}.
\end{equation}

Thus for each observation as well as having the radial velocity $v_R$ of the plasma, we also have a lower limit to the escape velocity, $v_{\rm esc}(R_{\rm max})$ from the upper limit $R_{\rm max}$ to the radius at which the absorption feature is formed. We show these values in Table~\ref{tableA}. There we have omitted the observations with ID 0673510801 (for which no entry is given in Kosec et al., Table 5) and 0783770701 (for which the error bars given in Kosec et al., Tables 3~\&~5 are too substantial).

\begin{table}
  \centering
  \begin{tabular} {|l|c|c|c|} \hline
    Obs ID & $v_R$ km/s & $v_{\rm esc}(R_{\rm max})$ km/s & $f = v_R \sqrt{2} / v_{\rm esc}$ \\ \hline
    0134120101 & 270 & 8732 & 0.043 \\
    0153950301 & 230 & 1764 & 0.184 \\
    0673510501 & 1000 & 1528  & 0.926 \\
    0673510601 & 360 & 4688 & 0.109 \\
    0673510901 & 600 & 536 &1.583 \\
    0783770501 & 550 & 627 &1.241 \\
    0783770601 & 450 & 816 & 0.780 \\ \hline 
  \end{tabular}
  \caption{Table summarising the relevant data adapted from \cite{Kosec:2020aa}. The columns are the Observation IDs (column 1), the measured radial velocities (column 2), the escape velocity from the estimated maximum launching radius $R_{\rm max}$ (column 3), and the ratio $f = v_R \sqrt{2} / v_{\rm esc}$ (column 4).} \label{tableA}
\end{table}

\begin{figure}[htb]
\begin{tikzpicture}
  \node (img1)  {\includegraphics[scale=0.39]{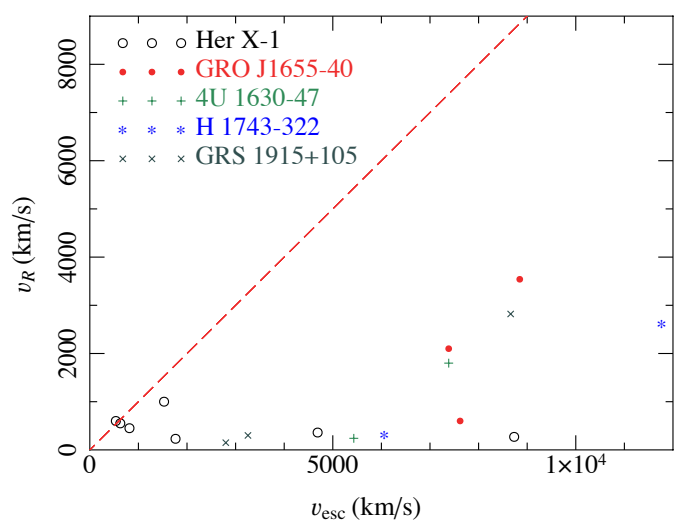}};
  \node[below of= img1, node distance=0cm, yshift=-3.5cm] {$v_{\rm esc}$\,(km/s)};
  \node[left of= img1, node distance=0cm, rotate=90, anchor=center,yshift=4.5cm] {$v_{R}$\,(km/s)};
\end{tikzpicture}
\caption{In this figure we plot the measured outflow radial velocity, $v_{R}$, against the lower limit to the escape velocity calculated from the estimated upper limit of the launching radius (see Tables 3 \& 5 of \citealt{Kosec:2020aa} for the data on Her X-1, and Tables 11 \& 12 of \citealt{Miller:2015aa} for the data on GRO J1655-40, 4U 1630-47, H 1743-322 \& GRS 1915+105). The error bars (not plotted) are substantial. The red dashed line describes $v_{R} = v_{\rm esc}$. In all but one case, the measured radial velocities are below (the lower limit to) the escape velocity.}
\label{fig1}
\end{figure}

\begin{figure}[htb]
\begin{tikzpicture}
  \node (img1)  {\includegraphics[scale=0.39]{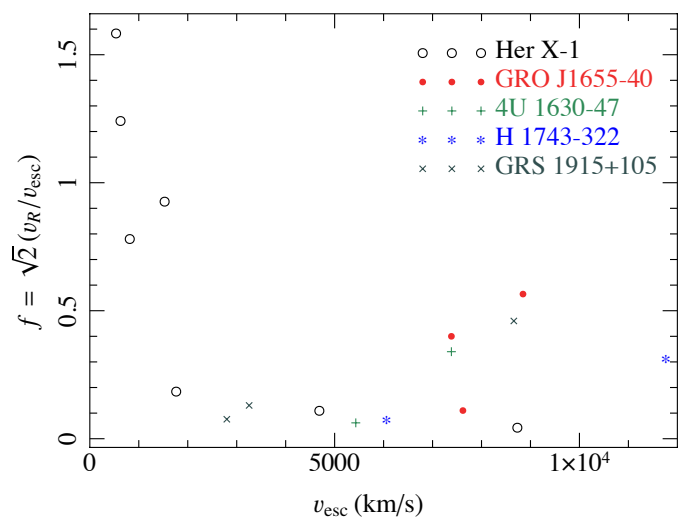}};
  \node[below of= img1, node distance=0cm, yshift=-3.5cm] {$v_{\rm esc}$\,(km/s)};
  \node[left of= img1, node distance=0cm, rotate=90, anchor=center,yshift=4.5cm] {$f = \sqrt{2}\left(v_{R}/v_{\rm esc}\right)$};
\end{tikzpicture}
\caption{In this figure we plot the value $f = v_R \sqrt{2} / v_{\rm esc}$ against the escape velocity for the data in Figure~\ref{fig1}. The error bars (not plotted) are substantial, and the estimated values of $f$ are all upper limits (see text). It is evident that most (if not all) observed values of $f$ lie below unity, indicating that the observed outflows are bound.}
\label{fig2}
\end{figure}

In Figure~\ref{fig1} we plot the observed radial velocity $v_R$ against the (lower limit to the) escape velocity from the estimated  radius at which the line is formed. We note that almost all of the radial velocities are lower than the escape velocity.  This is made clearer in Figure~\ref{fig2} where we plot the parameter $f$, defined in Section~\ref{intro}, again versus the estimated escape velocity at which the line is formed. Taking into account the large formal error bars which \cite{Kosec:2020aa} attach to their data, as well as other possible uncertainties it is evident that most of the outflowing material seen in Her X-1 is well below the escape velocity. 

Thus the blue-shifted emission lines observed by \cite{Kosec:2020aa} seem to indicate an outflow of material within a few tens of degrees of the surface of the disc whose velocities are (mostly) below the escape velocity. We see that  the estimated values of $f$ from these observations are in the right ballpark for the MCZ-driven outflows postulated by \cite{Nixon:2019ab}. We also note that the estimated outflow mass loss rates (Table~5 of \citealt{Kosec:2020aa}) are sizeable fractions of the estimated accretion rates, also in line with the requirements of the toy model for MCZ flows proposed by \cite{Nixon:2019ab}.

\section{Disc winds in other systems}
\label{discwinds}
Outflows of the kind reported by \cite{Kosec:2020aa} have also been reported in other systems, mainly in stellar mass black hole binaries \citep{Miller:2006aa,Miller:2006ab,King:2012aa,Ponti:2012aa,Trueba:2019aa}.  In all of these systems the disc surface is thought to close to the line of sight \citep{Miller:2006aa,Miller:2006ab,Ponti:2012aa}. A detailed analysis of four of these systems (4U 1630-47, GRO J1655-40, H 1732-322 and GRS 1915+105) is given by \cite{Miller:2015aa}. In these systems the analysis and modelling of the spectra requires the absorption lines to be modelled as being formed in two or three launching zones, depending on the source. \cite{Miller:2015aa} also deduce approximate radii for the formation of these lines using both photo-ionization models (as in \citealt{Kosec:2020aa}) and also using the width of accompanying emission components, assumed to correspond to the Keplerian velocity at the line-formation radius. Both these methods are in approximate agreement, although the uncertainties (error bars) are large. 

In Figure~\ref{fig1}, we plot the radial velocities $v_R$ measured for the absorption lines in these systems (given as $v_{\rm abs}$ in Table~11 of \cite{Miller:2015aa} against the escape velocity of the estimated formation radii (deduced from the values $r_{\rm phot}$ in Table~12 of \citealt{Miller:2015aa}). In Figure~\ref{fig2} we plot the corresponding values of the parameter $f$. In these objects it is evident that all of the observed gas is flowing outwards at well under the escape velocity. From Table~12 of \cite{Miller:2015aa} it is also evident that the mass fluxes involved in the outflows are all substantial fractions of the estimated accretion rates.

\section{Conclusions}
\label{conclusions}
We have investigated the properties of the so-called disc winds reported to be associated with the steady accretion discs in X-ray binaries, containing both black holes and neutron stars. The evidence for the wind or outflow is found in the form of absorption lines seen along the line of sight to the central source of X-rays -- being the disc centre in the black hole systems and the neutron star in Her X-1. There seems to be good evidence that these lines of sight pass close (typically around $10^\circ - 20^\circ$) to the disc surface \citep{Ponti:2012aa,Kosec:2020aa}. Analysis of the formation region of these lines is able to give an indication of the radii at which the lines are formed, as well as an estimate of the local gas densities. From these radii it is possible to estimate the escape velocities from the line formation regions. In Figures~\ref{fig1} and \ref{fig2} we have shown that the outflow velocities are typically significantly less than the escape velocities where the lines are formed. We conclude that these outflows are not unbound and so do not represent winds, in the conventional sense of the word. Rather they indicate the presence of outflows of material along the surface of the disc. Estimates of the energies involved in the outflows indicate that they involve a significant fraction of the accretion energy \citep{Miller:2015aa,Trueba:2019aa,Kosec:2020aa}.

We have argued that the properties of these disc surface outflows match the properties of the magnetically controlled outflows postulated by \cite{Nixon:2019ab} as an explanation of the anomalous properties of steady accretion discs seen in the nova-like variables, in terms of spatial structure, velocity structure and energy flux. \cite{Nixon:2019ab} postulated that such flows are driven by magnetic stresses originating in the disc -- in agreement with \cite{Miller:2015aa} who argue that in the absence of other possibilities the outflows that they observe must be magnetically driven.

If these conclusions are correct, we note that our our understanding of how (fully-ionized, steady) accretion discs operate needs to be substantially revised -- in particular with regard to the modelling of the magneto-rotational instability and resultant MHD turbulence, which is thought to drive such discs. It has already been noted  \citep{King:2007aa,Martin:2019aa} that numerical simulations of such discs are unable to match the high (viscosity parameter $\alpha \sim 0.3$) values of the disc viscosity required by the observations. We note further that although global numerical simulations of magnetized discs do appear to show surface flows, driven by magnetic stresses, the properties of such flows do not appear commensurate with what is observed. The general picture, resulting from several numerical investigations of strongly magnetised discs with substantial net vertical magnetic flux, is that winds ($v_R > v_{\rm esc}$) can be produced at vertical heights above the disc of $z \sim R$ with mass fluxes of order $1-10$ per cent of the accretion flow, and that surface flows involving several 10s of per cent of the accretion flow are present but always flow inwards \citep{Sadowski:2016aa,Zhu:2018aa,Jiang:2019aa,Zhu:2019aa,Mishra:2020aa}. So far it does not appear that these simulations have found the observed surface outflows with mass fluxes that are a significant fraction of the accretion flow and outward radial velocities that are of order the orbital velocity but significantly below the escape velocity.

\begin{acknowledgements}
We thank Mitch Begelman, Peter Kosec, Jean-Pierre Lasota and Jon Miller for sending comments on the manuscript, and the referee for a positive report. CJN is supported by the Science and Technology Facilities Council (grant number ST/M005917/1).
\end{acknowledgements}

\bibliographystyle{aasjournal}
\bibliography{nixon}




\end{document}